\documentclass[conference]{IEEEtran}
\IEEEoverridecommandlockouts
\usepackage{cite}
\usepackage{amsmath,amssymb,amsfonts}
\usepackage{algorithmic}
\usepackage{graphicx}
\usepackage{textcomp}
\usepackage{xcolor}
\def\BibTeX{{\rm B\kern-.05em{\sc i\kern-.025em b}\kern-.08em
    T\kern-.1667em\lower.7ex\hbox{E}\kern-.125emX}}
    
\begin{document}

\title{ Analyzing Photovoltaic's Impact on Conservation Voltage Reduction in Distribution Networks
\thanks{This work is supported by the National Science Foundation under ECCS 1929975.
}
}

\author{\IEEEauthorblockN{Rui~Cheng, Zhaoyu~Wang, Yifei~Guo, and Fankun~Bu
\IEEEauthorblockA{{Department of Electrical and Computer Engineering} \\
{Iowa State University}\\
Ames, IA, USA \\
ruicheng@iastate.edu, wzy@iastate.edu, yfguo.sdu@gmail.com, and fbu@iastate.edu}
}
}

\maketitle

\begin{abstract}

Conservation voltage reduction (CVR) has been widely implemented in distribution networks and helped utilities effectively reduce energy and peak load. However, the increasing penetration level of solar photovoltaic (PV) has affected voltage profiles and the performance of CVR. It remains an outstanding question how CVR and solar PV interact with each other. Understanding this interaction is important for utilities in implementing CVR and assessing its performance. This paper studies the impact of solar PV on CVR in a real distribution system in the Midwest U.S. using comprehensive simulations. We have considered various PV allocations and penetration levels, as well as different inverter control modes according to IEEE Std 1547-2018. Three metrics are used to quantify the impact of solar PV on CVR: voltages at the substation, voltage distribution across the network, and energy consumption reduction due to CVR. The results show that the allocations of solar PV have the most significant effect on the CVR performance, where a dispersed allocation of solar PV will help flatten voltage profile and achieve deeper voltage reductions at the substation, less energy consumption and line losses.
\end{abstract}


\section{Introduction} 
Conservation voltage reduction (CVR) is known as a cost-effective method to save energy and reduce peak demand \cite{zw,RWU}. By intentionally lowering voltages in distribution systems in a controlled manner, CVR can shave peak load and reduce energy consumption while keeping the lowest customer utilization voltage consistent with levels determined by regulatory agencies and standards-setting organizations \cite{DK}--\cite{ANSI}. According to the ANSI Standard C84.1\cite{ANSI}, service voltage must be at a minimum of 114 V, which is 5$\%$ below 120 V and utilization voltage. 

In general, CVR is accomplished by controlling Volt/VAr regulation devices such as capacitor banks, on-load tap changers (OLTCs), and voltage regulators \cite{RRJ}. For example, many utilities implement CVR by adjusting tap positions of substation OLTCs. The nature of CVR is that load is sensitive to voltage. Therefore, the performance of CVR depends on load compositions and voltage reduction depth. The power industry uses the CVR factor to quantify the performance, which is defined as the ratio between the percentage of load reduction and the percentage of voltage reduction. 

Recent years have seen a dramatic surge of solar photovoltaic (PV) in distribution networks. However, the increasing PV penetration can have significant influences on voltage profiles, thus affecting the CVR implementation. To achieve a better performance for the CVR implementation, it is of great importance to explore the interaction between CVR and solar PVs. Several works related to solar PVs and CVR have been done in previous studies. The study in \cite{ZW2} optimizes solar photovoltaic placement through Volt/VAr optimization (VVO) to minimize the total load consumption. The coordination between conventional voltage regulation devices, e.g., OLTCs, and solar inverters, are formulated as VVO problems in \cite{QZK}-\cite{MSH} to implement CVR in distribution networks. The simulation results in \cite{AB} indicate that a proper integration of solar PVs, operated at constant power factor mode, can contribute to alleviating the voltage violation caused by CVR, but the comprehensive relationship and analysis between CVR and solar PVs are not studied and discussed in detail. These works analyze the interaction between CVR and solar PVs from the optimization point of view based on the assumption that solar PV placement can be planned and PV inverters can be controlled in a centralized way. Nevertheless, in reality, solar PVs are installed by customers without a central optimization and operated locally in accordance with inverter control modes without a centralized optimization. In short, it is still an outstanding problem how CVR and solar PV interact with each other under different scenarios (e.g., different PV allocations).

\begin{figure*}[t]
    \centering
    \includegraphics[width=7in,height=4.6in]{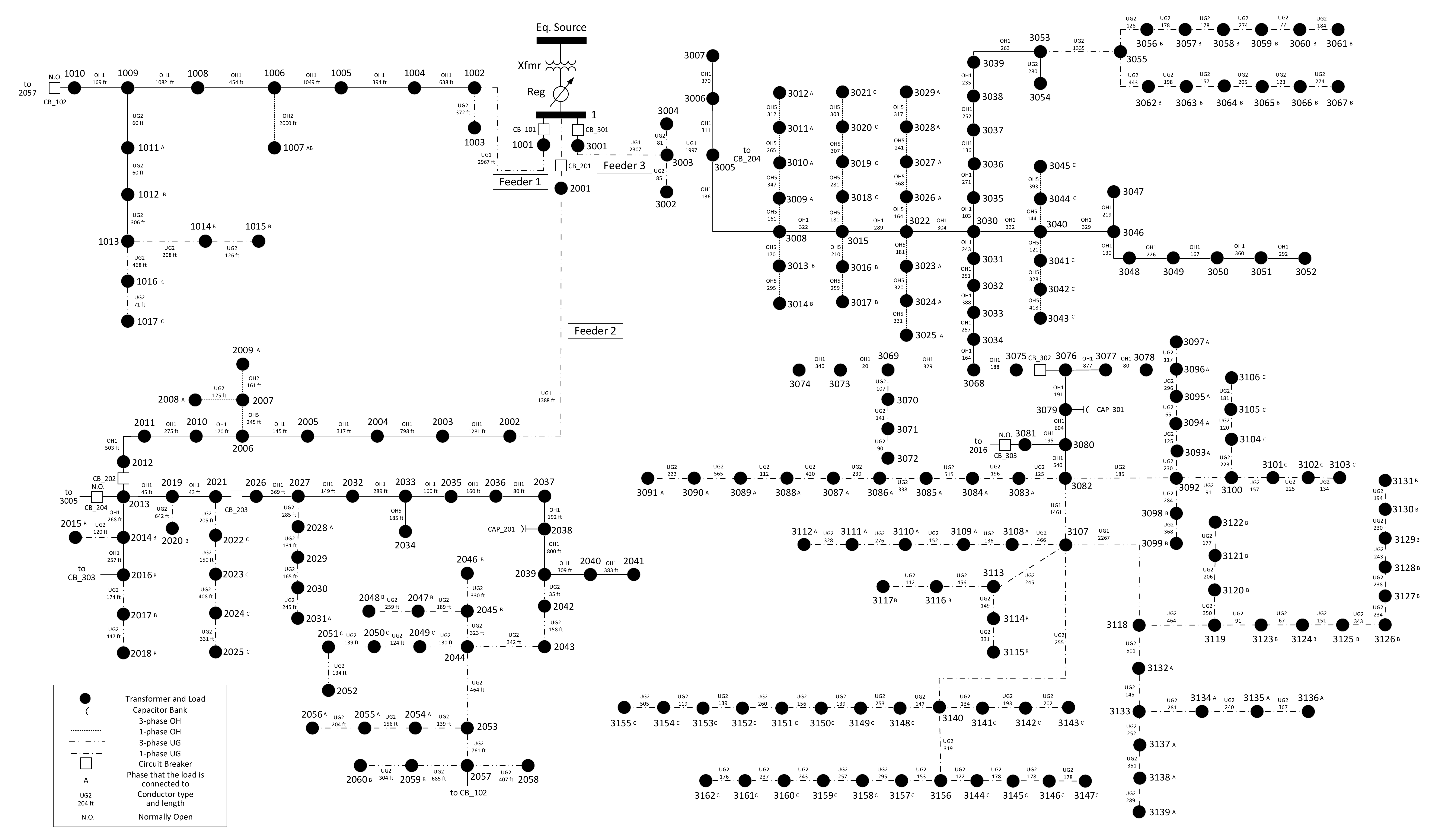}
    \caption{Topology of the distribution network \cite{FB}}
    \label{topology}
\end{figure*}

To bridge this gap, our paper provides a thorough analysis of the interaction between CVR and solar PV. All simulations are performed on a real distribution network located in the Midwest U.S. \cite{FB} with the open-source simulation package OpenDSS \cite{OpenDss}. The CVR is implemented by changing the tap position of the substation OLTC. Different PV penetration levels, allocations, and inverter control modes, including the constant power factor (PF) mode and voltage-reactive power (VRP) mode in IEEE Std 1547-2018 \cite{1547-2018}, are considered in our simulations. The simulation results are analyzed and discussed from three different metrics: voltages at the substation, voltage distribution across the network, and energy consumption reduction. The results show that the PV allocations have the most significant impact on the performance of CVR. As PVs are installed in a dispersed way, a more flattened voltage profile can be expected, leading to less energy consumption and line losses. Besides, the effect of inverter control modes on CVR is closely related to the PV allocations. The VRP mode can achieve a clear better performance in terms of energy saving and reducing line losses compared to the PF mode as PVs are installed in a dispersed way.





\section{Network Modeling}\label{data}
\subsection{Topology of the Network}
A real distribution test system \cite{FB} located in the Midwest U.S. is adopted in this study, as shown in Fig. \ref{topology}. It is a radial distribution network consisting of three feeders with 240 primary buses and 23 miles of conductors, which are connected to a 69/13.8 kV step-down  three-phase  substation  transformer. Note that customers are connected to these primary network nodes via secondary distribution transformers.
\subsection{ZIP Load Model}
\begin{figure}[t]
    \centering
    \includegraphics[width=3.5in]{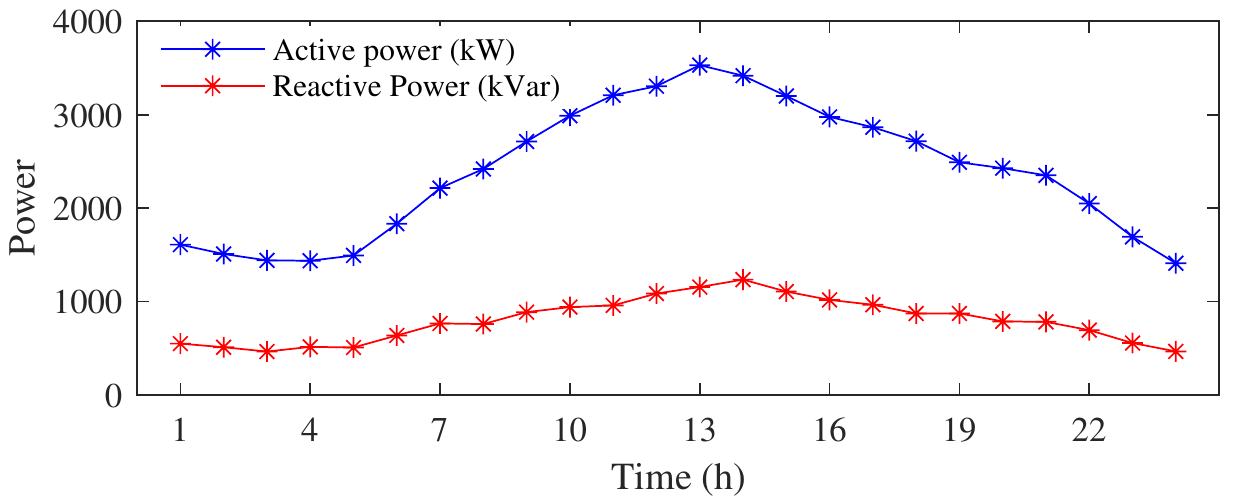}
    \caption{Time-series aggregate load across one day}
    \label{load}
\end{figure}

To simulate the CVR effect, all customers are modeled as voltage-sensitive loads. The loads connected to the distribution network are represented as a static load model with polynomial ZIP coefficients. The polynomial expressions for active and reactive power of ZIP models are
\begin{equation}
    P=P_{0}\left[Z_{p}(\frac{V_i}{V_0})^2+I_{p}(\frac{V_i}{V_0})+P_{p}\right],\, Z_{p}+I_{p}+P_{p}=1
\end{equation}
\begin{equation}
    Q=Q_{0}\left[Z_{q}(\frac{V_i}{V_0})^2+I_{q}(\frac{V_i}{V_0})+P_{q}\right],\, Z_{q}+I_{q}+P_{q}=1
\end{equation}
where $P$ and $Q$  are the actual active and reactive power at the operating voltage $V_{i}$, $P_{0}$ and $Q_{0}$ are the active and reactive power at the rated voltage $V_{0}$, $Z_{p}$, $I_{p}$ and $P_{p}$ are the ZIP coefficients of active power, $Z_{q}$, $I_{q}$ and $P_{q}$ are the ZIP coefficients of reactive power. In this work, we set $Z_{p}=Z_{q}=0.5, I_{p}=I_{q}=0.3, P_{p}=P_{q}=0.2$. The time-series aggregate load at the rated voltage across one day is shown in Fig. \ref{load}, where the time resolution is one hour.

\subsection{Solar PV and Inverter Models}
Solar PV units are connected to grids through inverters. The combination of PV and inverter model \cite{CR},\cite{PR} have been translated in OpenDSS input files. According to the guidance and requirement of reactive power capability of the Category B DER in IEEE Std 1547-2018 \cite{1547-2018}, we set properties of PV and inverter model \cite{CR} in OpenDss as $\%CutIn=5$, $\%CutOut=5$, $\%PminNoVars=5$, $\%PminkvarMax=20$ and $kvarMax$, $kvarMaxAbs$ and $kVA$ satisfy:
\begin{subequations}
\begin{align}
kvarMax&=0.44kVA\\
kvarMaxAbs&=0.44kVA
\end{align}
\end{subequations}
\begin{figure}[t]
    \centering
    \includegraphics[width=3.5in]{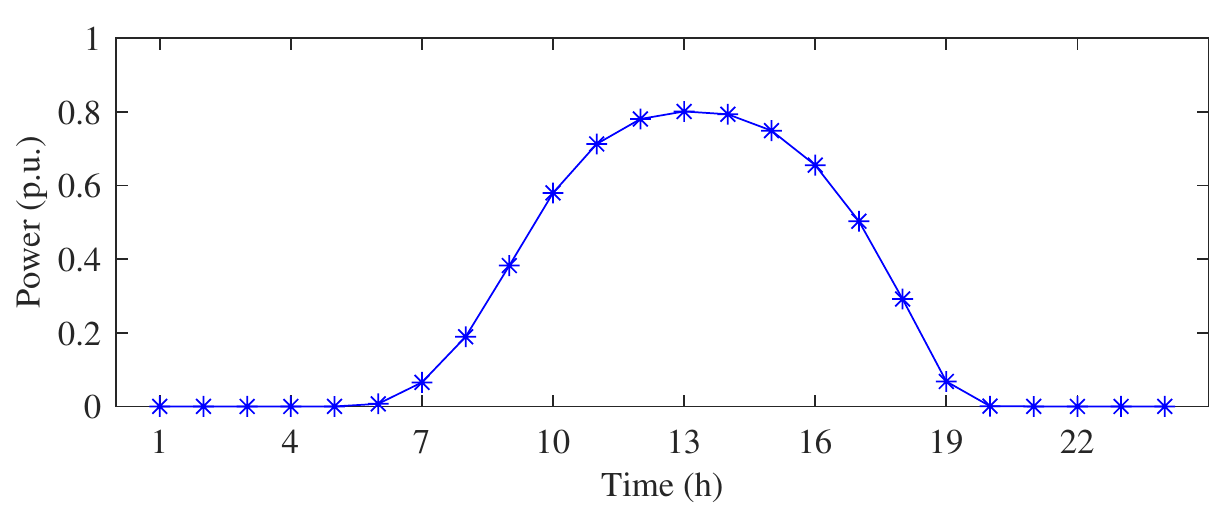}
    \caption{PV curve across one day}
    \label{PV}
\end{figure}

In this work, PV penetration is defined as the ratio of the aggregate peak power of PV units and the system peak load. A typical daily PV curve, shown in Fig. \ref{PV}, is used to simulate the change of PV power output (PV penetration can be adjusted by changing the peak power of PV units). Three different PV allocations are considered and referred to as Allocation 1, 2 and 3, respectively. Allocation 1 and 3 can be regarded as the aggregate PV allocation.

(1) Allocation 1 (Head): Several large PV units are installed at the head of feeders. The aggregate peak power of PV units is set as 60 $\%$ of the system peak load.

(2) Allocation 2 (Dispersed): PV units are installed in a dispersed way where each customer is equipped with one PV unit. We assume that the PV peak power of each customer is proportional to the customer's load at the system peak load hour. The peak power of each PV unit is set as 60$\%$ of its corresponding customer's load at the system peak load hour.

(3) Allocation 3 (End): This allocation consists of several large PV units installed at the end of the feeder. The aggregate peak power of large PV units is set as 60$\%$ of the system peak load.

Another important aspect to study is the inverter control mode. Two main inverter control modes in IEEE Std 1547-2018 \cite{1547-2018} are considered here: PF mode and VRP mode:

(1) PF mode: As for the constant PF mode, we directly set the power factor of PV as $PF=1$ in OpenDss. 

\begin{figure}[t]
    \centering
    \includegraphics[width=3.5in]{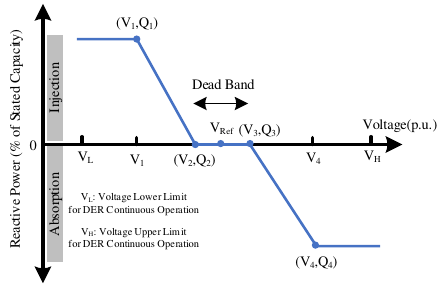}
    \caption{Voltage-reactive power characteristic\cite{1547-2018}}
    \label{VVCC} 
\end{figure}

(2) VRP mode: As for the voltage-reactive power mode, the PV units actively control their reactive power output as a function of voltage following a voltage-reactive power piece-wise linear characteristic, which is shown in Fig. \ref{VVCC}. According to the default parameter values specified in IEEE Std 1547-2018 \cite{1547-2018}, we set (V$_1$,Q$_1$)=(0.92,0.44), (V$_2$,Q$_2$)=(0.98,0), (V$_3$,Q$_3$)=(1.02,0), (V$_4$,Q$_4$)=(1.08,-0.44), V$_\text{Ref}$=1, V$_\text{L}$=0.9, V$_\text{H}$=1.1.

\subsection{OLTC Control Strategy}
Feeders are supplied by a 69/13.8 kV step-down three-phase substation transformer associated with three independent single-phase tap changers, each tap changer can be adjusted within a range from -16 (Lower) to 16 (Raise), corresponding to the voltage range from 0.9 p.u. to 1.1 p.u. In this paper, CVR is implemented by controlling the tap changers. To implement CVR while maintaining the reliability of the distribution system, the OLTC control strategy is implemented as follows:

(1) The OLTC operates in a phase-wise manner.

(2) To achieve the deepest voltage reduction by CVR, the voltage of OLTC should be adjusted as low as possible while satisfying network constraints, i.e., the voltages across the networks are in the range of [0.95, 1.05].

\section{Numerical Result Analysis}\label{sec:NRA}
\subsection{Base Case and Setting}
The distribution system without CVR and PV is used as the base case. When there is no CVR implementation, the substation voltage (i.e., OLTC voltage) is always fixed at 1 p.u. in this paper.
\begin{figure}[htb]
    \centering
    \includegraphics[width=3.3in]{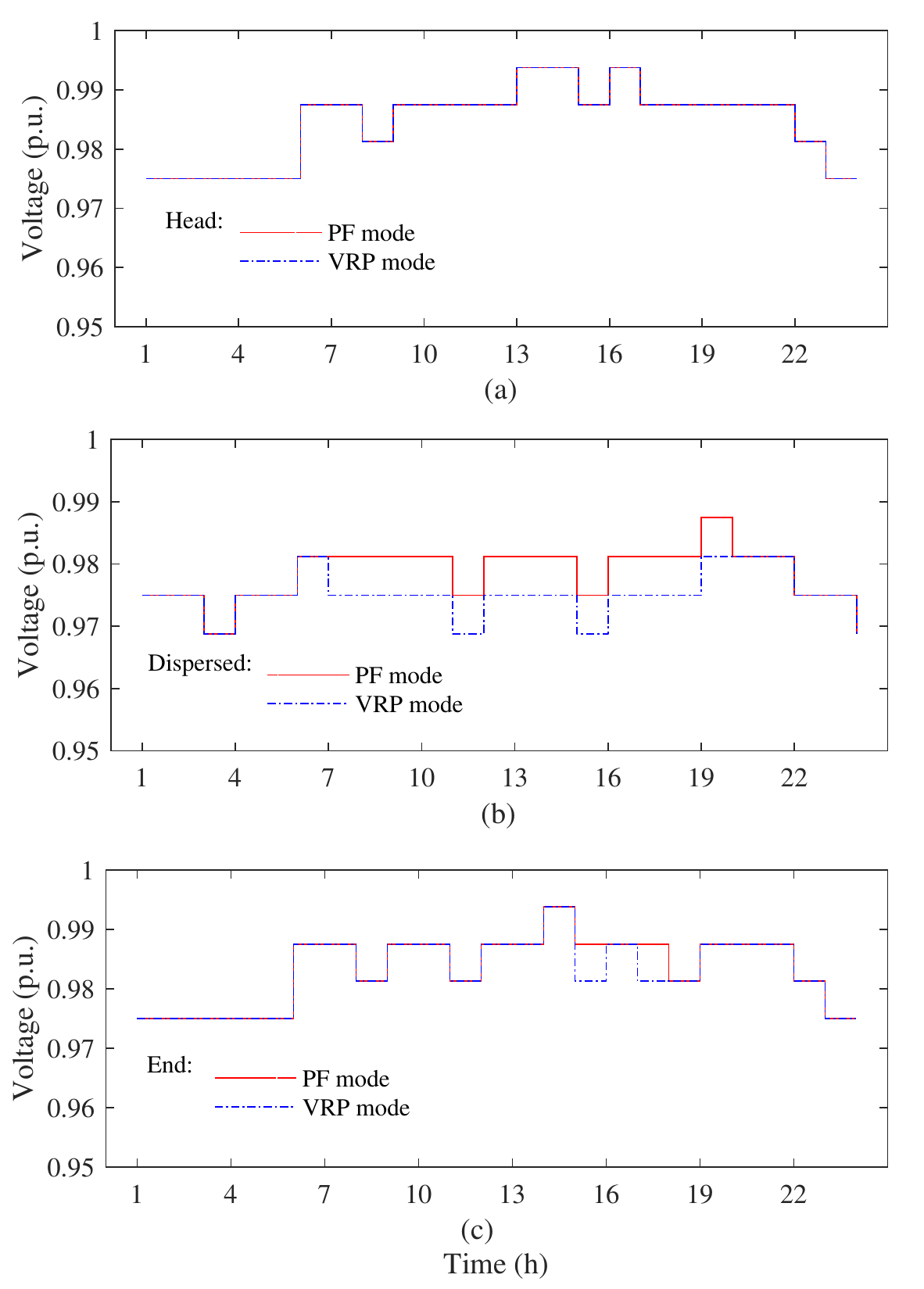}
    \caption{Phase A voltages at the substation under different scenarios: (a) Head; (b) Dispersed; (c) End}
    \label{OLTC}
\end{figure}
\subsection{Voltages at the Substation}
As mentioned in Section \ref{data}.D, CVR is implemented via OLTC. Thus, voltages at the substation are important to investigate the effect of solar PV on the CVR implementation. Taking phase A as an example, phase A voltages at the substation under different scenarios are shown in Fig. \ref{OLTC}.  There are two main observations from Fig. \ref{OLTC}:

(1) In general, phase A voltages at the substation for the VRP mode are not higher than the PF mode. Compared to the PF mode with $PF=1$, extra reactive power can be injected from solar inverters to the grid under the VRP mode as long as the inverters satisfy the operation constraints, which could lead to deeper voltage reductions at the substation. Note that the difference of voltage reduction between the VRP mode and the PF mode is clear as the PV units are installed in the Dispersed allocation. However, it is not the case with respect to the Head and End allocations. This is because the reactive power output of inverters in the VRP mode is highly related to the PV allocations. As the large PV units are installed at the head or the end of feeders, the voltage at the head or the end of feeders is relatively high, which impedes the reactive power generation.

(2) As a whole, phase A voltages at the substation for the Dispersed allocation are usually lower than the Head and End allocations. For the Head and End allocations, the customer loads are supplied from the energy resources at the head and end of feeders. However, with respect to the Dispersed allocation, each customer load can be supplied by its local solar PV. There is less power flow across feeders, which results in a more flattened voltage profile. The more flattened voltage profile could lead to deeper voltage reductions at the substation. 

\subsection{Voltage Distribution Across Network}
\begin{figure}[t]
    \centering
    \includegraphics[width=3.3in]{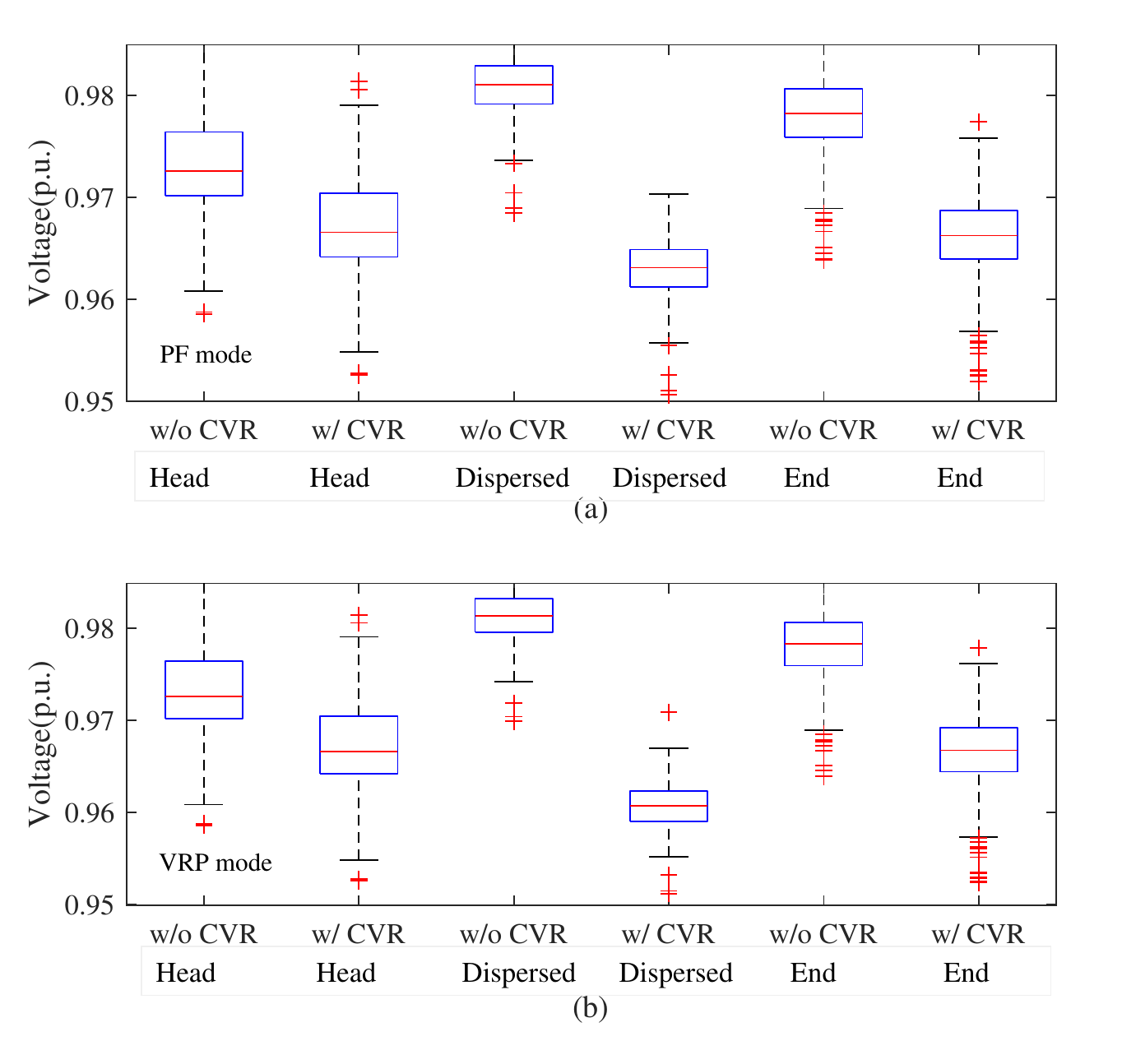}
    \caption{Voltage distribution across the distribution network at 13h: (a) PF mode; (b)VRP mode}
    \label{vd}
\end{figure}

Given the voltage-sensitive characteristics of ZIP load, the performance of CVR essentially depends on the voltage distribution across the distribution network.  Taking 13h as an example, Fig. \ref{vd} shows the voltage profiles with and without CVR across the distribution network at 13h. As shown in Fig. \ref{vd}, in both the PF and VRP modes, the overall voltages across the network for the Dispersed allocation are higher than the Head and End allocations when there is not CVR implementation. However, the overall voltages across the network for the Dispersed allocation are the lowest among the three allocations. With respect to the Dispersed allocation, there would be less power flow across the network since each customer load can be supplied by its local PV, leading to higher overall voltages compared to the Head and End allocations without CVR.  Meanwhile, the less power flow always contributes to a more flattened voltage profile. It facilitates deeper voltage reductions at the substation with CVR, which results in the lowest overall voltages after the CVR implementation. 

\begin{figure}[t]
    \centering
    \includegraphics[height=1.6in,width=3.0in]{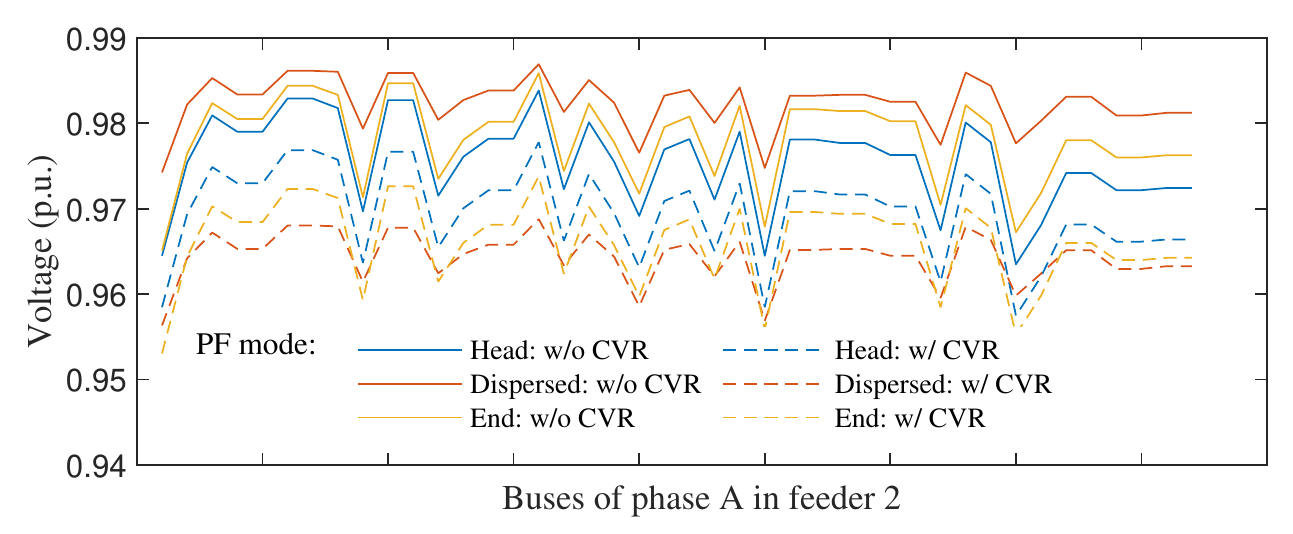}
    \caption{Phase A voltage profile of feeder 2 under PF mode at 13h}
    \label{vp}
\end{figure}

Fig. \ref{vp} shows the phase A voltage profile of feeder 2 under constant PF mode at 13h. As shown in Fig. \ref{vp}, the voltage profile without CVR for the Dispersed allocation is higher but more flattened compared to the Head and End allocations. And the voltage profile with CVR for the Dispersed allocation is lower and more flattened due to deeper voltage reductions at the substation. 

\begin{figure}[t]
    \centering
    \includegraphics[width=3.3in]{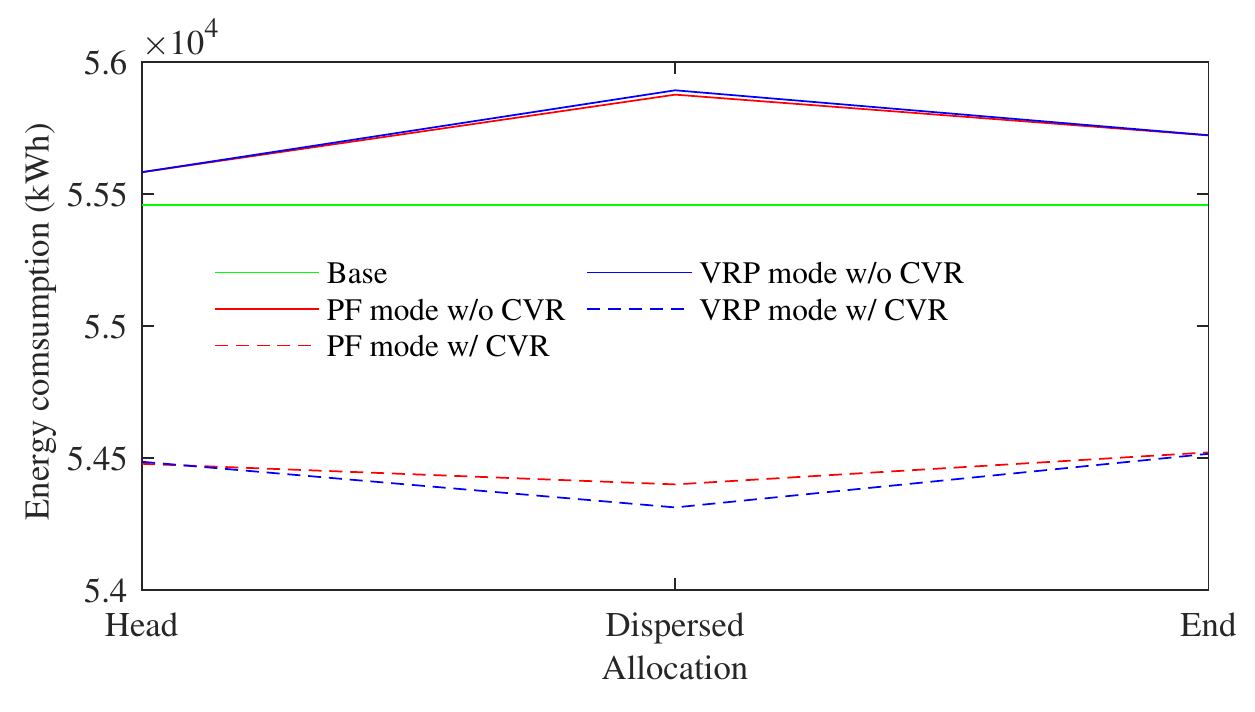}
    \caption{Total energy consumption under different allocations and modes}
    \label{tc}
\end{figure}
\begin{figure}[t]
    \centering
    \includegraphics[width=3.3in]{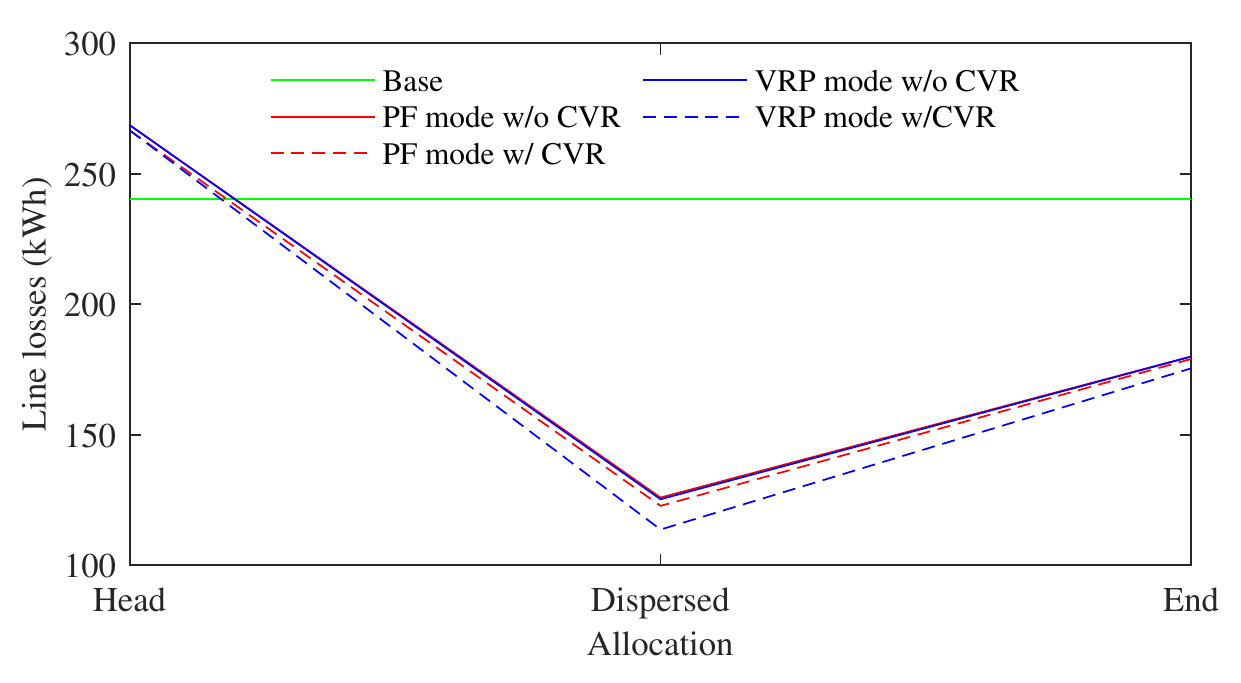}
    \caption{Line losses under different allocations and modes}
    \label{Lineloss}
\end{figure}

\subsection{Energy Consumption Reduction}
Fig. \ref{tc} and Fig. \ref{Lineloss} show the total energy consumption and line losses under different scenarios. Note that the total energy consumption is purely the energy consumed by customers, PV generation is not included in it. The total energy consumption under different allocations and modes is higher than the base case before the CVR implementation, but it is lower after the CVR implementation, indicating that the CVR is still an effective way to reduce energy consumption as PV units are connected to the grid. 

It can be observed from Fig. \ref{tc} that the total energy consumption under the Dispersed allocation are lowest after the CVR implementation. As discussed in Section \ref{sec:NRA}.C, the overall voltages across the network under the Dispersed allocation is lowest after the CVR implementation, leading to the lowest total energy consumption. In addition, as seen in Fig. \ref{Lineloss}, the line losses under the Dispersed allocation are lower than the Head and End allocations due to the less power flow across the network.

As for total energy consumption and line losses, the PF and VRP modes do not show a clear difference under the Head and End allocations, but the difference is clear under the Dispersed allocation. The VRP mode can achieve a clear better performance in terms of energy saving and reducing line losses compared to the PF mode under the Dispersed allocation. As pointed in Section \ref{sec:NRA}.B, it could hinder the reactive power generation from inverters under the Head and End allocations due to the high voltages at the head and end of feeders.

\subsection{PV Penetration Effect}
\begin{figure}[t]
    \centering
    \includegraphics[width=3.3in]{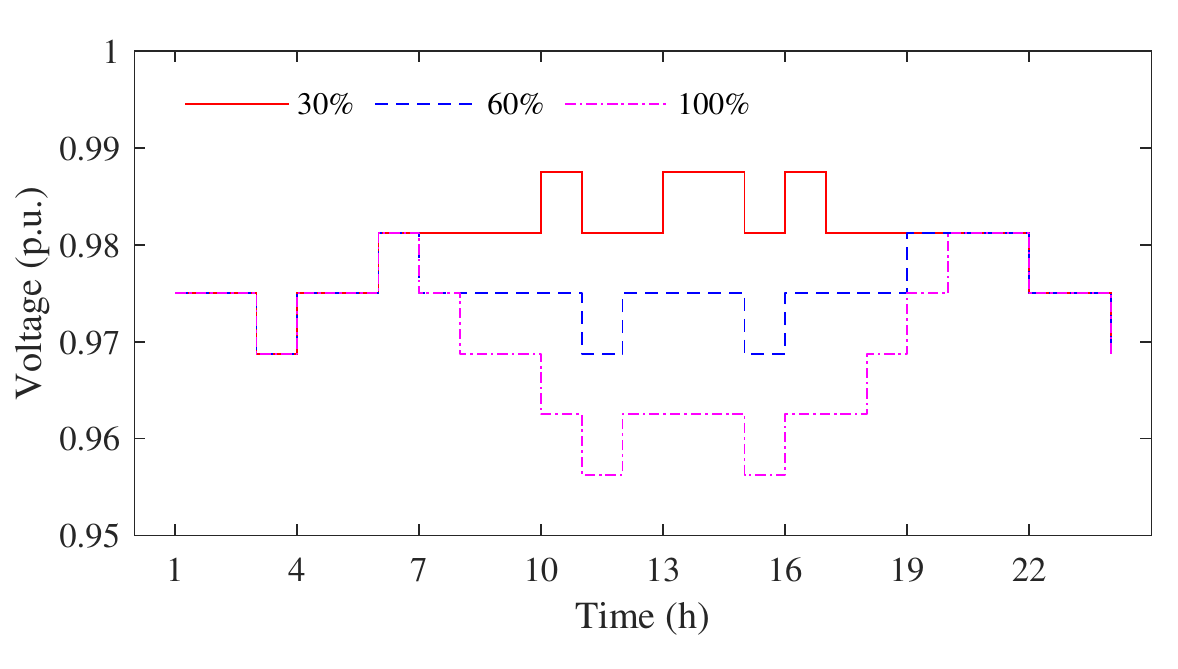}
    \caption{Phase A voltages at the substation under different PV penetration levels}
    \label{TapPenetration}
\end{figure}

To further investigate the effect of PV penetration on CVR, three different PV penetration levels are considered in this subsection: 30$\%$, 60$\%$, and 100$\%$ PV penetration levels by changing the peak power of PV units. PV units are assumed to be operated in the VRP mode and the Dispersed allocation. Fig.\ref{TapPenetration} shows phase A voltages at the substation under different PV penetration levels. It can be observed from Fig. \ref{TapPenetration} that phase A voltages at the substation decreases as the PV penetration increases. Higher PV penetration always leads to deeper voltage reductions at the substation for the CVR implementation. 

\begin{figure}[b]
    \centering
    \includegraphics[width=3.3in]{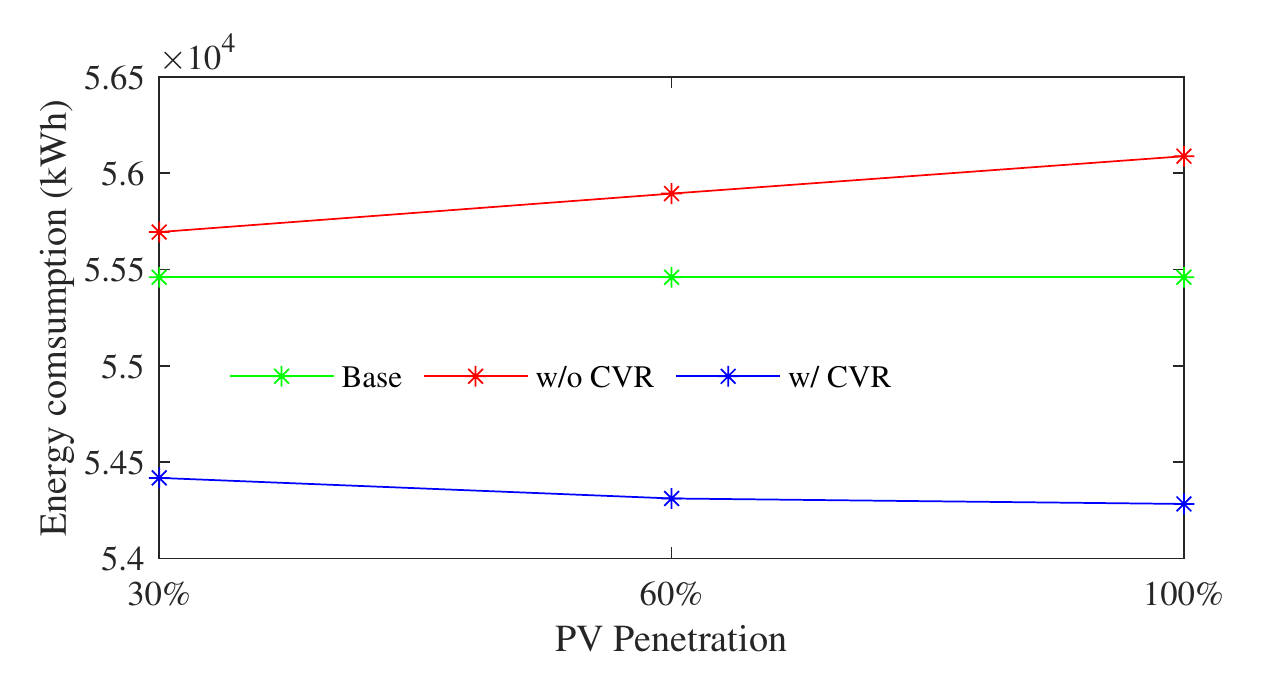}
    \caption{Total energy consumption across one day under different PV penetration levles}
    \label{Penetration}
\end{figure}
Fig. \ref{Penetration} shows the total energy consumption across one day under different PV penetration levels. As shown in Fig. \ref{Penetration}, total energy consumption without CVR increases as the PV penetration increases since the voltages across networks increase with the increasing PV penetration. However, total energy consumption does not show the same tendency after the CVR implementation. It indicates that the deeper voltage reductions at the substation by implementing CVR can offset the effect of increasing voltages on energy consumption, which is caused by the increasing PV penetration level.

\section*{Conclusions}
This paper studies the interaction between CVR and solar PV based on comprehensive simulations in a real distribution feeder in the Midwest U.S. with different PV allocations, control modes, and penetration levels. There are several interesting observations during the analysis. Compared with the aggregate allocations, i.e., the Head and End allocations, there are less energy consumption and line losses  under the Dispersed allocation due to more flattened voltage profiles. The effect of CVR implementation is closely related to the flattening degree of voltage profiles. The difference between the PF and VRP modes is not clear for the Head and End allocations, but it is not the same case for the Dispersed allocation. Besides, it is also observed that a higher PV penetration level always leads to deeper voltage reductions at the substation.

\end{document}